\documentclass[aps,twocolumn,groupedaddress,showpacs]{revtex4}
\usepackage{graphicx}
\usepackage{dcolumn}
\usepackage{bm}
\usepackage{epstopdf}
\usepackage[dvipsnames,usenames]{color}
\usepackage{color}
\usepackage{float}
\usepackage{amsmath}
\usepackage{amssymb}
\usepackage{amsthm}
\usepackage{mathrsfs}

\usepackage[T1]{fontenc}
\usepackage[skip=1ex]{caption}
\usepackage{booktabs,tabularx}
\usepackage{siunitx}

\renewcommand{\thefootnote}{*}
\emergencystretch=\maxdimen
\begin{document}
\title{Spontaneous-emission-enabled dynamics at the threshold of a directly modulated semiconductor laser}
\author{Junlong~Zou$^{1,3}$, Hanxu~Zhou$^{1,2}$, Can~Jiang$^{1,2}$, Gaofeng~Wang$^{1,2}$, Gian Luca~Lippi$^{4}$, and Tao~Wang$^{1,2}$}

\affiliation{$^1$Engineering Research Center of Smart Microsensors and Microsystems of MOE, Hangzhou Dianzi University, Hangzhou, 310018, China}
\affiliation{$^2$School of Electronics and Information, Hangzhou Dianzi University, Hangzhou, 310018, China}
\affiliation{$^3$School of Communication Engineering, Hangzhou Dianzi University, Hangzhou, 310018, China}
\affiliation{$^4$ Universit\'e C\^ote d'Azur, Institut de Physique de Nice (INPHYNI), UMR 7010 CNRS, 1361 Route des Lucioles, F-06560 Valbonne, France}

\date{\today}

\begin{abstract}
Chaos in semiconductor lasers or other optical systems has been intensively studied in the past two decades. However, modulation around threshold has received much less attention, in particular in gain-modulated semiconductor lasers. In this article, we investigate the bifurcation sequence which appears with pump modulation in the threshold region with a large amplitude and different values of modulation frequency. Modulation around threshold necessarily includes ``below-threshold'' dynamics, which can be effectively displayed only through through a nonlinear visualization of the oscillations. The irregular temporal behaviour is examined at various modulation frequencies and amplitudes, highlighting a possible route to chaos for very large amplitude modulation in the near-threshold region. The addition of the (average) spontaneous emission to the lasing mode enables a coupled dynamics between photons and carriers even below threshold, thus extending the pump range in which the modulation actively modifies the laser behaviour. We also report on the existence of a transition between similar attractors characterized by a temporal transient which depends on the amplitude of the modulation driving the pump.
\end{abstract}


\maketitle 

\section{Introduction}
The field of laser physics and chaos theory developed independently until 1975 when Haken discovered a striking analogy between the Lorenz equations that model fluid convection and the Maxwell-Bloch equations describing light-matter interaction in single mode lasers~\cite{Haken1975, Haken1978, Lugiato1985}. The following decades have seen a strong development in the study of laser dynamics, particularly under modulation of a control parameter, and in its applications~\cite{Sciamanna2015}. Single mode semiconductor lasers, as class B systems~\cite{Tredicce1985}, possess an intrinsically two-dimensional dynamics and transition into chaotic oscillations once an external parameter is modulated~\cite{Ohtsubo2007}.  

While an outstandingly large body of work has been conducted on macroscopic lasers, the emergence of nanoscale devices opens perspectives for new phenomena and applications~\cite{Hill2014,Deng2021}, thanks not only to their reduced footprint and energy consumption, but also to their different stability properties~\cite{Wang2021} and the peculiarities of their phase space~\cite{Wang2020JMO,Lippi2021AM} which remain for the most part unexplored.  The combination of larger noise fraction, gradual transition from incoherent to coherent emission~\cite{Wang2015} and increased dynamical stability may result in  different patterns of instabilities leading to unforeseen functional uses.

General dynamical features have been identified in nanolasers~\cite{Lorke2011, Hachair2011, Lebreton2013, Moelbjerg2013, Pan2016, Wang2017, Wang2018, Puccioni2018, Rasmussen2019, Wang2019JOSM}, together with dynamics induced by feedback~\cite{Otto2010, Otto2012, Lingnau2013, Munnelly2017, Holzinger2018a, Holzinger2018b, Holzinger2018c, Wang2019JSTQE, Wang2020PRA, Rasmussen2021OE}, mode competition~\cite{Leymann2013, Redlich2016, Schlottmann2019, Schmidt2021}, mode coupling and synchronization~\cite{Kreinberg2019}.  Sensing~\cite{Wang2021sensor} and generation of low-coherence light~\cite{Wang2021nanomat} have been recently proposed as a result of the dynamical features which are typical of these devices.  

A more systematic approach to the study of the dynamics at small scales requires the establishment of a connection between macroscopic and nanothe- devices.  Since the latter differ mostly in their behaviour around the threshold region, a regime which has been largely neglected in macroscopic lasers, it is important to establish the dynamical response of a macroscopic laser in this same pump regime when its energy supply is being modulated.  We therefore choose to investigate the dynamics of a single-mode macroscopic semiconductor laser modulated across threshold with a large modulation amplitude using a modelling approach which includes the (very small) contribution of the spontaneous emission to the lasing mode (cf. Section~\ref{whySE} for the motivations).  This latter contribution must be included here, since it holds a crucial role in the dynamics of nanolasers.  The choice of large amplitude modulation reflects the need to match the needs of small scale devices~\cite{Wang2019JOSM}, where the broad extent of the threshold region requires a substantial amount of current modulation to sample the two emission regimes dominated by spontaneous and stimulated emission, respectively.  In order to set the stage for future comparison, in this paper we focus on this parameter region concentrating on the purely deterministic aspects of the problem: this will enable us to prepare a cartography of the dynamical response, to be compared to the equivalent one emerging from smaller lasers. The addition of noise -- of paramount importance for nanolasers -- will represent an additional step, to be considered once the deterministic features have been understood.

The manuscript is organized as follows: after a brief presentation of a standard model for single-mode semiconductor lasers, we examine two different aspects of modulation, (1) response to a frequency scan at fixed amplitude, and (2) response to an amplitude scan at fixed frequency. The fixed parameter is chosen to match a high-sensitivity region which allows for the observation of a variety of dynamical responses. We remark that the choice of the modulation-free bias point -- slightly above threshold -- does not fundamentally alter the observed dynamics. Hence, we can consider the results as representative of a typical laser response.  

As the ratio between the photon flux above and below threshold in a macroscopic laser is very large, the dynamics evolves over several decades of laser output. For this reason we have employed a logarithmic representation of the phase space to distinguish the attractors. This somewhat unusual choice may call into question its possible experimental implementation. While suitable schemes could be conceived to match the challenge, we stress that this property is peculiar to macroscopic lasers, since in nanolasers the dynamical range can decrease to one or two orders of magnitude. In addition, it is worth remarking that, since the model is sufficiently generic to possess features common to general Rate Equations~\cite{Siegman1986}, which satisfactorily describe solid state lasers, a possible experimental implementation of our scheme could be envisaged in those devices. There, the modulation frequency shifts from the GHz to the MHz range, thus enabling the use of logarithmically amplified, very sensitive detectors, capable of experimentally reproducing the features of the logarithmically displayed attractors.

\section{Rate equations}\label{REs}
The single mode rate equations for the carrier density $N$ and photon density $S$ of a semiconductor laser can be described by~\cite{Liu1993}:

\begin{eqnarray}
\label{modeqN}
\frac{dN}{dt} = \frac{I_{tot}}{qV} - \frac{N}{\tau_n} - G(N - N_0)\frac{S}{1 + \varepsilon S}\, , \\
\label{modeqS}
\frac{dS}{dt} = \Gamma G(N - N_0)\frac{S}{1 + \varepsilon S} - \frac{S}{\tau_s} + \Gamma \beta B N^2
\end{eqnarray} 

\noindent where $N$ represents the carrier and $S$ the photon densities, respectively. On the right hand side of eq. (\ref{modeqN}), the first term represents the modulated injection current, $I_{tot} = I_{dc} + I_msin(2\pi f_m t)$, where $I_{dc}$ and $I_m$ are the bias current and modulation amplitude, respectively, and $f_m$ is the modulation frequency; $q$ is the electric charge, and $V$ is the active volume. The second term is the carrier spontaneous decay with a lifetime $\tau_n$. The last term describes the stimulated emission of radiation where $G$ indicates the gain coefficient, and $N_0$ is the carrier density at transparency. In order to fit more realistic laser characteristics, the nonlinear gain compression (factor $\varepsilon$) is introduced for better agreement with experimental results~\cite{Hemery1990}. In eq. (\ref{modeqS}), the $\Gamma$ parameter represents the optical confinement factor, $\tau_s$ is the photon lifetime, and $\beta$ denotes the spontaneous emission coupling emission factor, which is the fraction of the spontaneous emission coupled into the lasing mode~\cite{Wang2015, Deng2021}. It is this last term which introduces the average contribution of the spontaneous emission to the photon number, thus enabling the establishment of deterministic trajectories across the lasing threshold. In principle, the carrier lifetime is a function of $N$~\cite{Selberherr1984} for which we use the following functional form: 

\begin{eqnarray}
\frac{1}{\tau_n} = A + BN + CN^2
\end{eqnarray} 

\noindent where A is the recombination coefficient, B the bimolecular recombination coefficient, and C the Auger recombination coefficient. Detailed parameter values are summarized in Table~\ref{tab1}. It is worth remembering that laser modulation through the pump requires much larger amplitudes than modulation through other parameters, such as cavity losses~\cite{Tredicce1985B}.

\begin{table*}[h!t]
\begin{center}
\setlength{\tabcolsep}{5mm}{
\caption{List of the laser parameters and their typical values, used in the simulations.}
\label{tab1}

\begin{tabular}{c c c c}
\toprule
\textbf{Parameter}	& \textbf{Symbol}	& \textbf{Value} & \textbf{Unit}\\
\midrule
Active volume		& $V$			& 6.75$\times$ 10$^{-11}$   & cm$^3$\\
Gain coefficient		& $G$			& 1.15$\times$10$^{-6}$  & cm$^3$s$^{-1}$\\
Electric charge		& $q$			& 1.6$\times$10$^{-19}$  & C\\
Carrier density at transparency & $N_0$			& 1.10$\times$10$^{18}$  & cm$^{-3}$\\
Gain compression factor		& $\varepsilon$			& 1.00$\times$10$^{-17}$  & cm$^3$\\
Optical confinement factor		& $\Gamma$			& 0.35  & -\\
Photon lifetime		& $\tau_s$			& 1.75$\times$10$^{-12}$  & s\\
Spontaneous emission factor		& $\beta$			& 1.00$\times$10$^{-7}$  & -\\
Recombination coefficient		& $A$			& 1.00$\times$10$^{8}$  & s$^{-1}$\\
Bimolecular recombination coefficient		& $B$			& 1.25$\times$10$^{-10}$  & cm$^3$s$^{-1}$\\
Auger recombination coefficient		& $C$			& 3.50$\times$10$^{-29}$  & cm$^6$s$^{-1}$\\
\bottomrule
\end{tabular}
}
\end{center}
\end{table*}

\subsection{Role of spontaneous emission}\label{whySE}

Normally neglected when modelling macroscopic lasers, spontaneous emission plays a substantial role in micro- and nanolasers. Thus, in order to establish a bridge between the wealth of information gathered on large lasers and their smaller counterparts, we need to see what is its influence even when only a small amount of photons is added to the dynamics.

Spontaneous emission is traditionally introduced in the rate equations through a proportionality coefficient -- called $\beta$-factor -- which represents the average number of incoherent photons (or photon density, in normalized variables) added to the lasing mode~\cite{Coldren2012}. The contribution of the spontaneous fraction adds to the lasing photons, introducing an important conceptual modification to the model.  If the background (noisy field) is neglected, once the carrier number is driven below the lasing threshold the coupling between the two dynamical variables (photons and carriers) gradually disappears due to their exponential decay (with their respective time constants).  This implies that the carriers will asymptotically follow the dynamics imposed by the driving with their exponential time constant (low-pass filtering) until threshold is attained again.  

When instead the spontaneous emission is reintroduced -- even if its contribution is small as in macroscopic lasers -- the coupling between the two variables does not disappear and the dynamics persists even ``below threshold'', thanks to the small contribution to the photon number coming from the spontaneously relaxing carriers.  The change is conceptually important since it maintains a dynamics and extends -- even though in a peculiar way -- the attractors below what is a strict threshold in the idealized macroscopic device.  In other words, the introduction of the spontaneous emission through the $\beta$ factor enables a dynamical laser response even in the range where the macroscopic laser, in the {\it thermodynamic limit} (i.e., size tending to infinity), would just show a trivial decay towards the off-state. Notice that the timescale on which the coupled dynamics evolves is no longer the one which describes the simple exponential relaxation of the spontaneous-emission-free approach.

It is important to remark that the effect, for macroscopic devices, is most relevant if the modulation takes place across threshold since the smallness of the $\beta$ factor renders the influence of the spontaneous contribution negligible if the laser is modulated sufficiently far above threshold. Even though investigations of pump-modulated macroscopic lasers with the rate equations of Section~\ref{REs} have been carried out, this has been systematically done well above threshold; thus, the regime that holds promise at the nanoscale has been, so far, ignored and prompts our current investigation.

\section{Results and discussions}
Throughout the paper, the model is integrated using a fourth-order Runge-Kutta scheme using the ode45 integration routine in MatLab. The timestep used in the integration is $t_s = 0.1 ps$, while each point in parameter space is computed for a time interval $t_d = 1 \mu s$. The results do not change if $t_s$ is reduced:  the transitions remain the same. If instead $t_s$ is increased a degradation in the trajectories' quality and changes in the transitions appears.  This originates from the well-known requirement applying to numerical integration schemes for which $t_s$ needs to be at least one order of magnitude smaller than the fastest time constant present in the model (in our case $t_s \lesssim \frac{\tau_s}{10}$).

\subsection{Free running laser characterization}
\begin{figure}[!t]
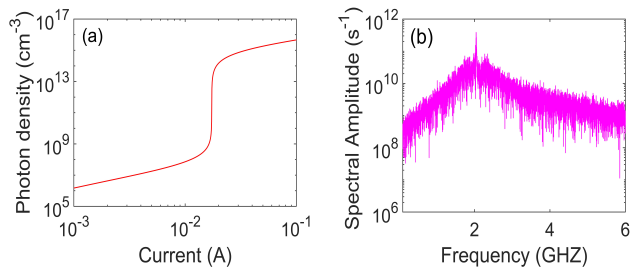

\centering
  \includegraphics[width=1.68in,height=1.40in]{I_O.pdf}
  \includegraphics[width=1.68in,height=1.40in]{rf.pdf}
  \caption{L-L function curve in log scale (a) and typical RF spectrum (b) of free running laser obtained at 1.14$I_{th}$.}
  \label{S-rf}
\end{figure}

We first investigate the emission properties of semiconductor laser under free running operation. Fig.~\ref{S-rf}a shows the laser's emission as a function of injection current in double-logarithmic scale (also traditionally called ``S-curve'' because of its shape). This curve is constructed by integrating the model for fixed injection current values, then taking the average photon number and plotting the ensemble of points obtained by varying the current.  This laser response exhibits a clear ``kink'' at $I_{th}$ = 0.019 A ($I_{th}$ indicates the laser threshold) and a  narrow transition between spontaneous and stimulated emission, indicating the typical threshold feature of low-$\beta$ lasers. Fig.~\ref{S-rf}b displays a typical radio frequency (RF) spectrum of the free running laser obtained at 1.14I$_{th}$, which shows a broad relaxation oscillation peak centered around 2.00 GHz. We therefore expect the most interesting dynamics to appear in a frequency range around this value, as we see below. It is important to notice that direct modulation in this frequency range is not trivial in prepackaged devices. This requires compensation of the parasitic imaginary contribution to the input impedance (stray capacitance and inductance). While doable, the problem can be avoided by using semiconductors on a die and high-frequency probes.

\subsection{Modulation frequency dependence}
\begin{figure}[!t]
\centering
  \includegraphics[width=2.6in,height=2.0in]{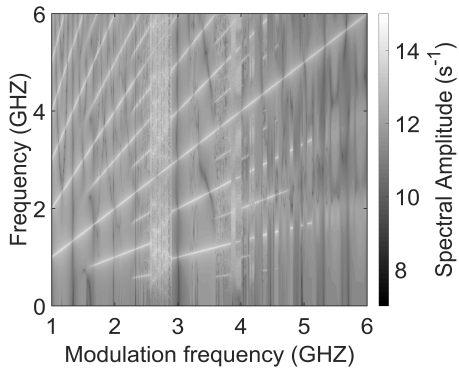}
  \caption{RF spectral mapping of semiconductor laser with fixed bias current ($I_{dc}$ = 1.14$I_{th}$), modulation amplitude ($I_m$ = 1.10$I_{th}$), as a function of modulation frequency.  The colour bar encodes the computed spectral amplitude of the laser frequency spectrum (as third dimension).}
  \label{various_modu_fre}
\end{figure}

We first analyze the laser response to the modulation frequency for fixed amplitude and bias current. Fig.~\ref{various_modu_fre} shows the RF spectral map at $I_{dc} = 1.14 I_{th}$ as a function of modulation frequency for a modulation amplitude $I_m = 1.10 I_{th}$. The higher-order harmonics, generated when $f_m < 1.65$GHz, are clearly visible. Period doubling distinctly appears for 1.65 GHz $< f_m <$ 2.30 GHz, followed by a narrow period quadrupling and period eight regions when the modulation frequency in the range of 2.30 GHz $< f_m <$ 2.45 GHz and 2.45 GHz $< f_m <$ 2.50 GHz, respectively. In the range 2.55 GHz $< f_m <$ 2.90 GHz, the whole RF spectrum broadens, suggesting chaotic oscillation. The broad spectral region is followed by a period 3 and then 6 response -- consistently with well-known, general bifurcation routes~\cite{Arecchi1982,Tredicce1985B,Tredicce1986}   in modulated class B lasers --, succeeded by more transitions to oscillations with wide spectral features (3.85 GHz $< f_m <$ 3.91 GHz and 4.14 GHz $< f_m <$ 4.22 GHz). Finally, the spectrum indicates that the laser follows the sinusoidal modulation when $f_m >$ 5.00 GHz.

In the following, we are going to concentrate on the range of modulation frequencies which contain the more complex dynamics, ranging up to approximately 2.8 GHz.  For faster modulations, Fig.~\ref{various_modu_fre} shows a less complex dynamics characterized by the alternance between a period-doubled behaviour and a direct following of the modulation, with repeats at spectral harmonics originating from the deformation of the sinusoidal driving.  The less complex behaviour is consistent with the RF spectrum of Fig.~\ref{S-rf}b.

\begin{figure*}[!t]
\centering
  \includegraphics[width=7.0in, height=4.5in]{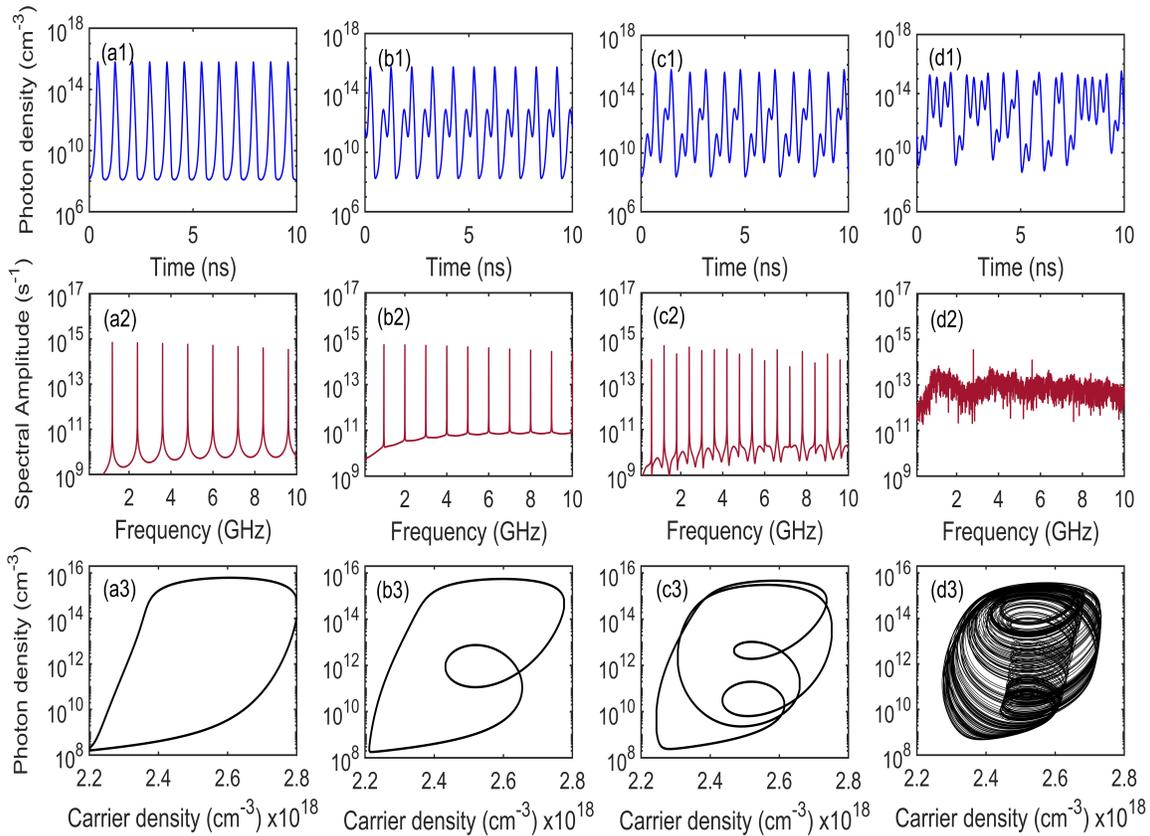}
  \caption{Temporal dynamics (log-linear), RF spectra and phase space trajectories as function of modulation frequency for a semiconductor laser biased at $I_{dc} = 1.14I_{th}$, and modulated with amplitude $I_m = 1.10I_{th}$ and $f_m$: 1.1 GHz (a1-a3); 2.0 GHz (b1-b3); 2.4 GHz (c1-c3); 2.8GHz (d1-d3).}
  \label{Frequency}
\end{figure*}

Fig.~\ref{Frequency} shows the temporal dynamics, RF spectra and phase portraits for laser modulation at $I_{dc} = 1.14I_{th}$ and $f_m$ = 1.1, 2.0, 2.4 and 2.8 GHz, respectively. The large modulation amplitude ($I_m = 1.10I_{th}$) enables gain-switching operation: the laser is periodically driven far below threshold, with the generation of regular, sharp pulses (Fig.~\ref{Frequency}a1) recognizable by the higher order harmonics of the modulation frequency (Fig.~\ref{Frequency}a2). The RF frequency spectrum (Fig.~\ref{various_modu_fre}) shows a peculiar feature of this choice of modulation whose dynamical response presents an approximate symmetry relative to the diagonal (laser output oscillation vs. current modulation). Its origin rests on the large amplitude modulation around threshold which carries with it the generation of sharp pulses (Fig.~\ref{Frequency}), thus an approximate repetition of the dynamical spectral features at higher harmonics. Physically, the large amplitude driving extends the attractor well below threshold, where memory is maintained through the tiny contribution of the average spontaneous emission ($\Gamma \beta B N^2$ term in eq. (2)). In spite of the smallness of this term, due to the value of $\beta$ in a macroscopic laser, it enables the system to maintain a deterministic dynamical trajectory even though the photon number is very small. This is the reason why a phase space portrait based on a logarithmic representation (cf. later) is most suited and capable of identifying the trajectories, as opposed to the traditional linear representation in photon number.  It also represents a peculiarity of the laser system perturbed by large amplitude modulation around threshold, a point which has, so far, not been broadly studied.


At the resonance frequency ($f_m \approx 2 GHz$, close to the laser relaxation oscillations~\cite{Coldren2012}), the pulse amplitude is enhanced (Fig.~\ref{Frequency}b1), in a similar way to what observed in a mesoscale laser~\cite{Wang2019JOSM}. Here, we remark on the appearance of a spectral subharmonic component (at 1.0 GHz, Fig.~\ref{Frequency}b2), consistent with previous observations~\cite{Lee1985}. Period-doubling is confirmed by the phase portrait (Fig.~\ref{Frequency}b3 -- logarithmic vertical scale), where a double-loop is observed. The pulse quality degrades when increasing the modulation frequency and at $f_m = 2.4 GHz$ a period quadrupling bifurcation occurs (Fig.~\ref{Frequency}c2). Due to the fact that most of the dynamics takes place at very low photon numbers, the phase portrait presents a quadruple-loop structure (Fig.~\ref{Frequency}c3). Finally, when the modulation frequency reaches 2.8 GHz, the temporal pulses appear to become random and their amplitude irregular (Fig.~\ref{Frequency}d1). The spectrum broadens (Fig.~\ref{Frequency}d2) and the phase space is compatible with a temporal behaviour of a chaotic nature. These simulations show that a period-doubling structure appears when the driving frequency is close to the above-threshold relaxation oscillation even when large amplitude driving brings the laser far below threshold, in transient. 

\subsection{Modulation amplitude dependence} 
We now examine the laser response to variations in modulation amplitude. Anticipating on the content of this subsection, we establish two interesting findings.  On the one hand, as observed in the modulation frequency sweep, a cascade of bifurcations is observed as a function of modulation amplitude. The phenomenon is not entirely trivial (as for the frequency modulation), since the laser is modulated across threshold with a very large modulation amplitude: the system is therefore driven very far below threshold for approximately half the cycle and only the (tiny) contribution of the average spontaneous emission enables the survival of trajectories (at extremely low photon numbers). On the other hand, we find a very unusual transition between attractors, which we characterize in some detail, at the beginning of the bifurcation cascade. No final conclusions are drawn on this observation and further work is needed to better understand the attractor features in this part of the phase space.
 
\begin{figure}[!t]
\centering
  \includegraphics[width=3.5in]{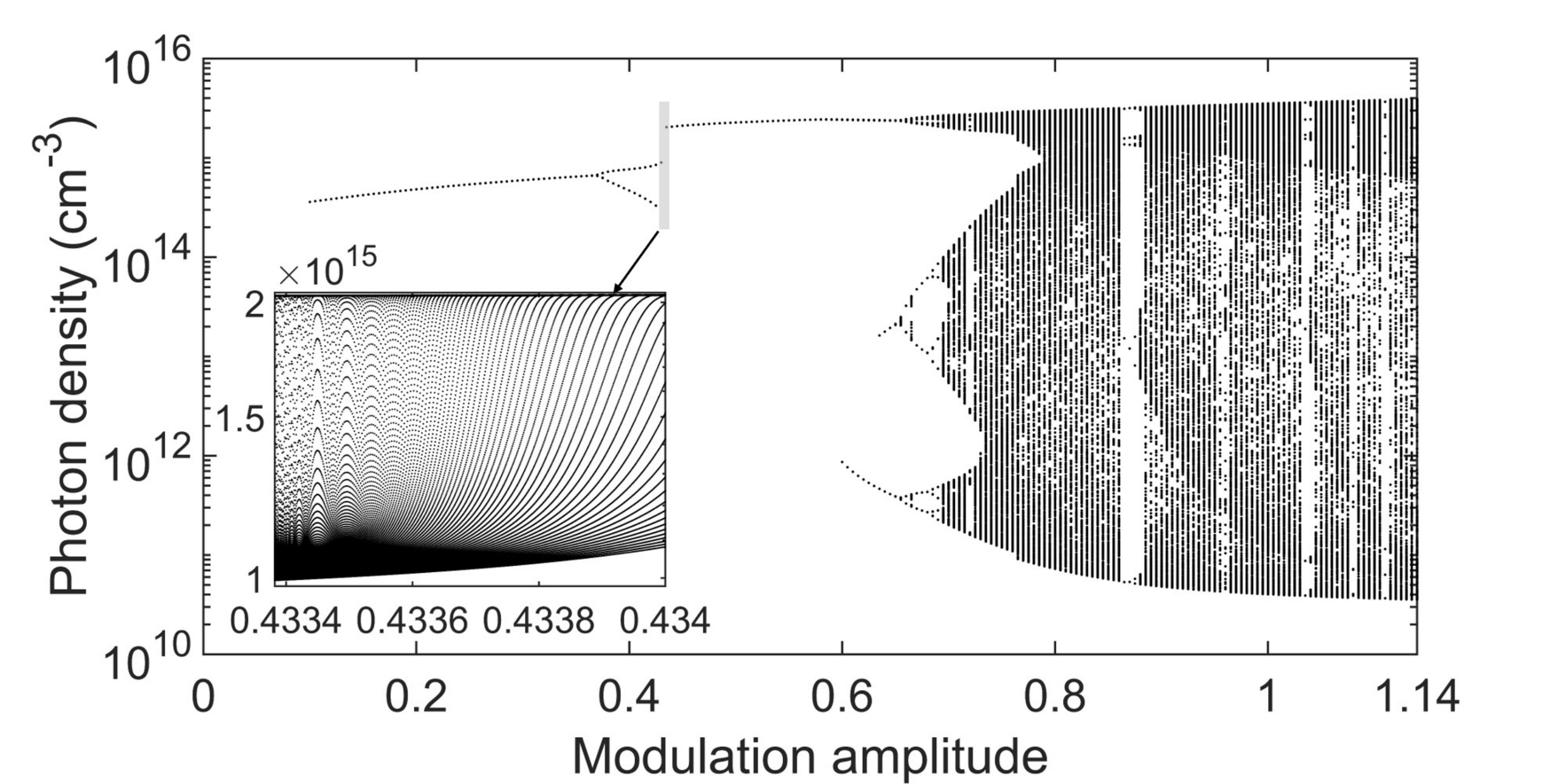}
  \caption{Bifurcation diagram as function of modulation amplitude for a semiconductor laser under 2.8 GHz modulation frequency.}
  \label{bifur-log}
\end{figure}

Fig.~\ref{bifur-log} shows the results of a scan in modulation amplitude ($0 \le I_m \le I_{dc}$ -- vertical axis in log scale) obtained by sampling the peak values of temporal pulses for the laser modulated at $f_m$ = 2.8 GHz and $I_{dc} = 1.14 I_{th}$. The diagram shows a complex structure as a function of $f_m$. Some of the finer features of doubling become apparent using the logarithmic representation used here. For $I_m \lesssim 0.37 I_{th}$ the laser follows the external modulation, to then present a period doubling bifurcation which evolves in the shaded area into a complex structure, which we examine in detail in the following (cf. inset). A sustained limit cycle of period $\frac{2}{f_m}$ and amplitude nearly constant persists until $I_m \approx 0.58 I_{th}$, where a further bifurcation appears, followed by a cascade similar to the one observed as a function of modulation frequency (Figs.~\ref{various_modu_fre} and~\ref{Frequency}).   

\begin{figure}[!t]
\centering
  \includegraphics[width=3.3in,height=2.80in]{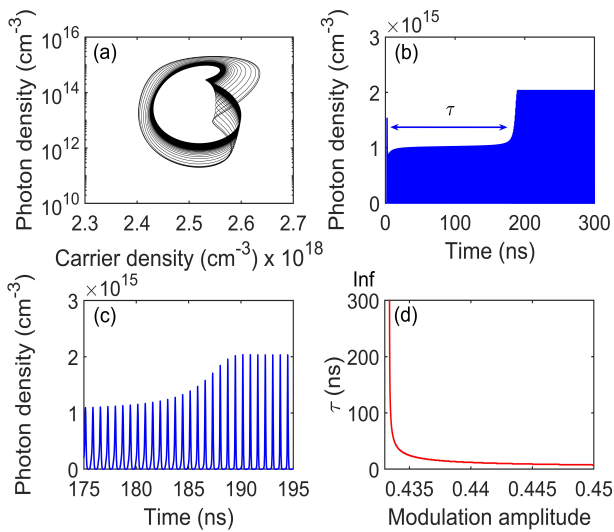}
  \caption{Phase space trajectories (a), temporal dynamics within 300 ns (b) and 20 ns (c) duration for modulation amplitude $I_m = $0.4334$I_{th}$; (d) time duration of transient pulses as a function of modulation amplitude.  $f_m = 2.8 GHz$ and $I_{dc} = 1.14 I_{th}$.}
  \label{transition}
\end{figure}

The shaded area highlights the parameter region which presents unusual topological features.  Starting from the period-doubled solution (double branch present for $I_m > 0.37 I_{th}$) a complex evolution of the amplitude takes place, illustrated in the inset of Fig.~\ref{bifur-log}. A high-resolution scan shows the succession of peak heights which develops a steady growth until the higher branch (main figure) is attained.  Further details can be gathered from the phase space picture and temporal evolution of the oscillation (Fig.~\ref{transition}).


\begin{figure}[!t]
\centering
  \includegraphics[width=3.4in,height=2.80in]{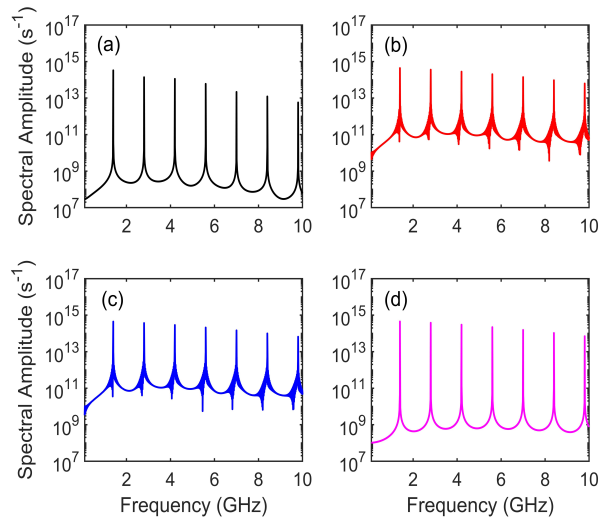}
  \caption{RF spectra calculated by Fourier transform of 2.7 million datapoints for the semiconductor laser modulated with amplitude $I_m$: (a) 0.43$I_{th}$; (b) 0.4334$I_{th}$, (c) 0.43342$I_{th}$, and (d) 0.435$I_{th}$, respectively.  $f_m = 2.8 GHz$ and $I_{dc} = 1.14 I_{th}$.}
  \label{RFspectra}
\end{figure}

As clearly illustrated by the temporal evolution of the signal at one of the amplitude levels ($I_m = 0.4334 I_{th}$, Fig.~\ref{transition}b), the Poincar\'e sections shown in the inset of Fig.~\ref{bifur-log} display the evolution of the transient which leads from a first, smaller attractor (inner one in Fig.~\ref{transition}a) towards a larger attractor (outer trajectory in the phase space representation, Fig.~\ref{transition}a). In the example shown, the transient takes more than 500 oscillation cycles to reach the larger trajectory, which has evolved from the smaller one. The transition between the two takes place without hysteresis, thus suggesting the disappearance of the first in favor of the second attractor. Fig.~\ref{transition}c shows a detail of the transition between the two types of oscillation: the smaller one with a broader pedestal and the larger one consisting of sharper and larger peaks. Both oscillations take place at $\frac{f_m}{2}$, as shown by the RF spectra of Fig.~\ref{RFspectra}.  Specifically, Fig.~\ref{RFspectra}a and~\ref{RFspectra}d show the spectra for the two different attractors. The one with smaller amplitude (Fig.~\ref{RFspectra}a) presents alternating heights for the frequency components, with a slight preference for the bifurcated ones (odd orders) and a clear reduction in the amplitudes of the harmonics (even the minima drop). The larger amplitude attractor (Fig.~~\ref{RFspectra}d) instead is characterized by a remarkable closeness between the $f_m$ and $\frac{f_m}{2}$ components and a much less pronounced amplitude decrease for the higher harmonics. The transient, presented in panels (b) and (c) for two slightly different values of $I_m$, possesses more complex spectral features which reflect the transition between the two kinds of oscillation (inset of Fig.~\ref{bifur-log} for comparison). The last interesting feature of this unusual attractor transition emerges from the dependence of duration of the transient $\tau$ as a function of $I_m$ (Fig.~\ref{transition}d). A sharp increase in the transient duration occurs as the modulation amplitude approaches the value for which the smaller attractor is stable, even though no scaling law appears to be identifiable.


Fig.~\ref{Dynamics-1} displays the typical temporal dynamics, RF spectra and phase space trajectories for modulation amplitudes $I_m =$ 0.20 $I_{th}$, 0.50 $I_{th}$, 0.60 $I_{th}$ and 0.677 $I_{th}$, respectively. In specific, for $I_m = 0.35 I_{th}$ regular, small amplitude pulses appear (Fig.\ref{Dynamics-1}a1) at the modulation frequency producing the limit cycle of Fig.\ref{Dynamics-1}a3. The response is, however, nonlinear as highlighted by the RF spectrum (Fig.\ref{Dynamics-1}a2). When $I_m = 0.50 I_{th}$, a significantly enhanced pulsing dynamics (Fig.\ref{Dynamics-1}b1) with period doubling (Fig~\ref{Dynamics-1}b2) emerges. In phase space, we find a closed loop containing a second loop (Fig.\ref{Dynamics-1}b3). At $I_m = 0.60I_{th}$ the laser is driven further below threshold, generating stronger and sharper pulses (Fig.\ref{Dynamics-1}c1) and presenting an additional subharmonic component ($T/4$, Fig.\ref{Dynamics-1}c2) and a double-looped phase space portrait (Fig.\ref{Dynamics-1}c3). Further increase in modulation amplitude ($I_m = 0.677_{th}$) extends the emission modifications to the peak heights (Fig.\ref{Dynamics-1}d1) with the appearance of an additional bifurcation (Fig.\ref{Dynamics-1}d2) and four similar structures in the phase portrait (Fig.\ref{Dynamics-1}d3).

\begin{figure*}[!t]
\centering
  \includegraphics[width=7.0in, height=4.5in]{Dynamics-1.pdf}
  \caption{Temporal dynamics (log-linear), RF spectra and phase space trajectories as function of modulation amplitude (from $I_m = 0.20I_{th}$ to 0.677$I_{th}$) for a semiconductor laser biased at $I_{dc} = 1.14I_{th}$ under 2.8 GHz modulation frequency.  $I_m$: 0.20$I_{th}$ (a1-a3); 0.50$I_{th}$ (b1-b3); 0.60$I_{th}$ (c1-c3); 0.677$I_{th}$ (d1-d3).}
  \label{Dynamics-1}
\end{figure*}

We now examine the details of a region in which the transition towards a regime compatible with chaotic emission takes place (cf. Fig.~\ref{bifur-log}). For $I_m = 0.695 I_{th}$ the pulse amplitude sequence starts to become more complex, with a partial decrease of the overall amplitude (Fig.~\ref{Dynamics-2}a1), maintaining a high subharmonic and harmonic content (Fig.~\ref{Dynamics-2}a2). The phase space shows trajectories which separate, substantially running along one another (Fig.~\ref{Dynamics-2}a3). A slight increase in modulation amplitude ($I_m = 0.712 I_{th}$) enhances the low amplitude components (Fig.~\ref{Dynamics-2}b1) with no substantial modifications of the RF spectrum (Fig.~\ref{Dynamics-2}b2) or of the phase space portrait, the sole difference being the thickness of the trajectory ensemble (Fig.~\ref{Dynamics-2}b3). When $I_m$ reaches to $0.75I_{th}$, small peaks become more obvious (Fig.~\ref{Dynamics-2}c1), but the subharmonic content starts to disappear, as shown by the RF spectrum (Fig.~\ref{Dynamics-2}c2). The phase space shows the occupation of a substantial volume by similar trajectories (Fig.~\ref{Dynamics-2}c3). At $I_m = 1.10 I_{th}$, higher amplitude, random-looking pulses take place (Fig.~\ref{Dynamics-2}d1) accompanied by a broad RF spectrum (Fig.~\ref{Dynamics-2}d2) and a dense phase space filling (Fig.~\ref{Dynamics-2}d3), compatible with a (nearly) featureless, broadband chaotic emission.  The overall evolution is consistent with an irregular emission, which at smaller modulation amplitude contains the traces of deterministic trajectories whose influence gradually disappears.  In other words, the further away the laser is driven from threshold, i.e. deeper into the spontaneous emission regime, the weaker the traces of a regular oscillation (even superposed to chaotic-like dynamics).  This kind of behaviour is rendered possible by the spontaneous emission contribution, since at such extreme modulation amplitude models for very large lasers (matching the {\it thermodynamic limit}, i.e., absence of spontaneous emission~\cite{Siegman1986}) would simply give an exponential relaxation of the photon number with little or no residual coupling to the carriers.


\begin{figure*}[!t]
\centering
  \includegraphics[width=7.0in, height=4.5in]{Dynamics-2.pdf}
  \caption{Temporal dynamics (log-linear), RF spectra and phase portraits as function of modulation amplitude (from $I_m = 0.695 I_{th}$ to $1.10 I_{th}$) for a semiconductor laser biased at $I_{dc} = 1.14I_{th}$ under 2.8 GHz modulation frequency.  $I_m$: 0.695$I_{th}$ (a1-a3); $0.712 I_{th}$ (b1-b3); $0.75 I_{th}$ (c1-c3); $1.10 I_{th}$ (d1-d2).}
  \label{Dynamics-2}
\end{figure*}

\section{Conclusions}
In conclusion, we presented the existence of a bifurcation structure which appears for a semiconductor laser for across-threshold, large amplitude pump modulation. The features are similar to those which emerge from above-threshold simulations, but for the use of a logarithmic representation, allowing for a clear trajectory identification at very small photon numbers. In a traditional linear scale phase portrait the below-threshold features would merge together blurring the picture and preventing the identification of distinct attractors. The existence of such trajectories when the laser is driven far below threshold comes from the inclusion of the average contribution of the spontaneous emission which, in concomitance with the memory effects imprinted into the carrier density, permits the prolongation of average, deterministic trajectories even when the laser is (practically) not emitting. In reality, a small quantity of photons is always present in the cavity, as reflected by the current results.  Though small, the latter is sufficient to introduce a degree of differentiation in the irregular, chaotic-like emission when the modulation amplitude is changed from more than half to nearly the value of the bias (maximum permitted change).

We have also identified the existence of a somewhat unusual transition between attractors of different amplitude but otherwise similar properties, characterized by different phase portraits and RF spectra.  The transition takes place as a continuous, bistability-free, temporal transient whose duration is controlled by the amplitude of the sinusoidal component modulating the pump.

\section*{Acknowledgment}
This work has been supported by the National Natural Science Foundation of China (Grant No. 61804036), Zhejiang Province Commonweal Project (Grant No. LGJ20A040001), National Key R \& D Program Grant (Grant No. 2018YFE0120000), Zhejiang Provincial Key Research \& Development Project Grant (Grant No. 2019C04003).

\end{document}